\documentclass[aps,showpacs,twocolumn,floatfix]{revtex4}
\usepackage{graphicx}
\usepackage{psfrag}
\usepackage{amsmath}

\begin{document}

\title{Square lattice site percolation thresholds for complex neighbourhoods}

\author{Mariusz Majewski}
\affiliation{Faculty of Physics and Applied Computer Science, AGH University of Science and Technology, al. Mickiewicza 30, PL-30059 Krak\'ow, Poland.}
\author{Krzysztof Malarz}
\homepage{http://home.agh.edu.pl/malarz/}
\email{malarz@agh.edu.pl}
\affiliation{Faculty of Physics and Applied Computer Science, AGH University of Science and Technology, al. Mickiewicza 30, PL-30059 Krak\'ow, Poland.}

\date{\today}

%% ----------------------------------------------------------------------------
\begin{abstract}
In this paper we compute the square lattice random sites percolation thresholds in case when sites from the 4th and the 5th coordination shells are included for neighbourhood.
The obtained results support earlier claims, that (a) the coordination number and the space dimension are insufficient for building universal formulae for percolation thresholds and (b) that percolation threshold may not decrease monotonically with lattice site coordination number.
\end{abstract}
%% ----------------------------------------------------------------------------

\pacs{
64.60 %% Percolation (phase transitions)
}

\maketitle

%% ############################################################################
\section{Introduction}
%% ############################################################################

The percolation \cite{ancient,sykes,percolation} is one of central topic of statistical physics with many applications from theory of the fluid flow through the porous media \cite{porous} to sociophysics \cite{socio} via immunology \cite{immunology,epi-net}, fires spread \cite{fires}, and many others \cite{percolation}.

Probably in all these fields the {\em percolation threshold} $p_c$ plays the crucial role:
When the sites occupation probability $p$ is larger than this critical value $p_c$ the liquids may flow from one side of porous rock to another, epidemic or disease may spread over all individuals of a population, and fire may destroy all trees in a model forest, etc. 
Thus evaluation of the percolation threshold for many networks \cite{epi-net,network} and lattices \cite{dd,lattice,pc-kagome,pc-3-12-12,pc-sl} and even surfaces \cite{surface} still appeals to scientists \cite{new} since the beginning of percolation theory \cite{ancient}. 
As exact values of percolation threshold for the site percolation are known only for several lattices 
--- for example:
1D chain of sites ($p_c=1$) \cite{percolation},
the triangular lattice ($p_c=1/2$) \cite{percolation},
Kagom\'e lattice ($p_c=1-2\sin(\pi/18)$) \cite{pc-kagome} 
and Archimedean $(3,12^2)$ lattice ($p_c=\sqrt{1-2\sin(\pi/18)}$) \cite{pc-3-12-12}
---
the main stream of investigation is computational (for analytical approach please refer to \cite{anal}).
For example, for square lattice with the nearest neighbours interaction the numerical estimation of $p_c(\text{2N})$ is $\theta=0.5927460(5)$ \cite{pc-sl}.

\begin{figure*}[!htbp]
\includegraphics[scale=0.5]{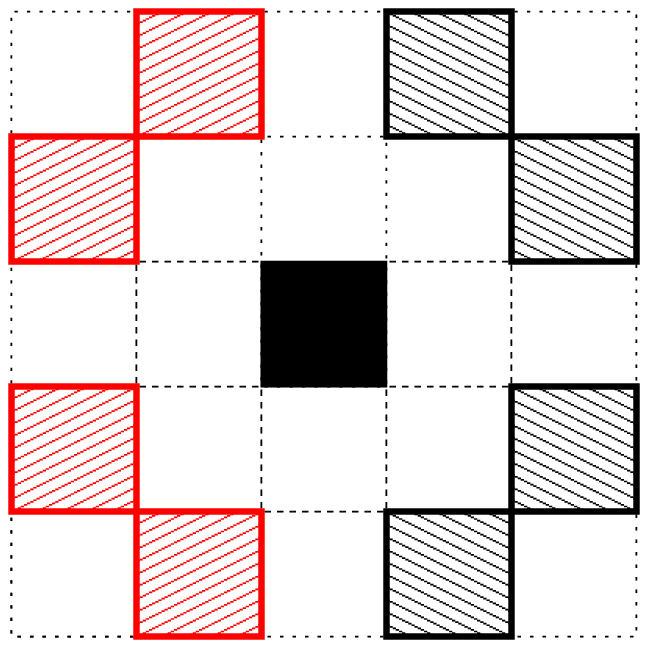}
\includegraphics[scale=0.5]{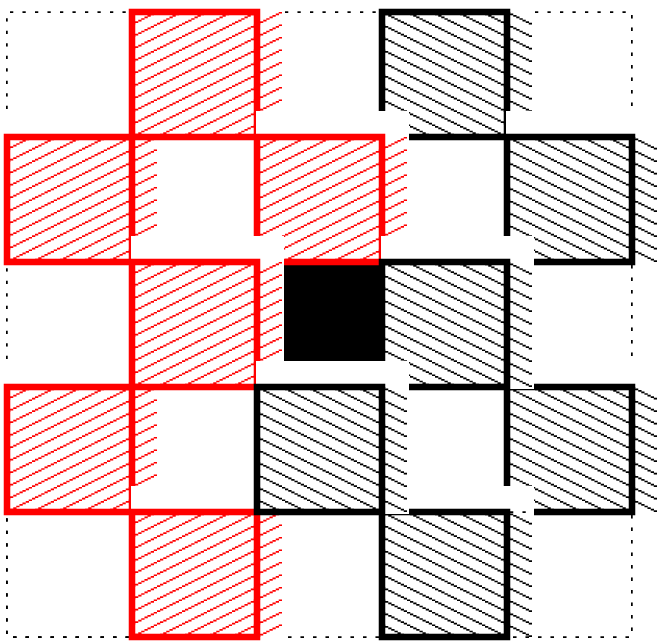}
\includegraphics[scale=0.5]{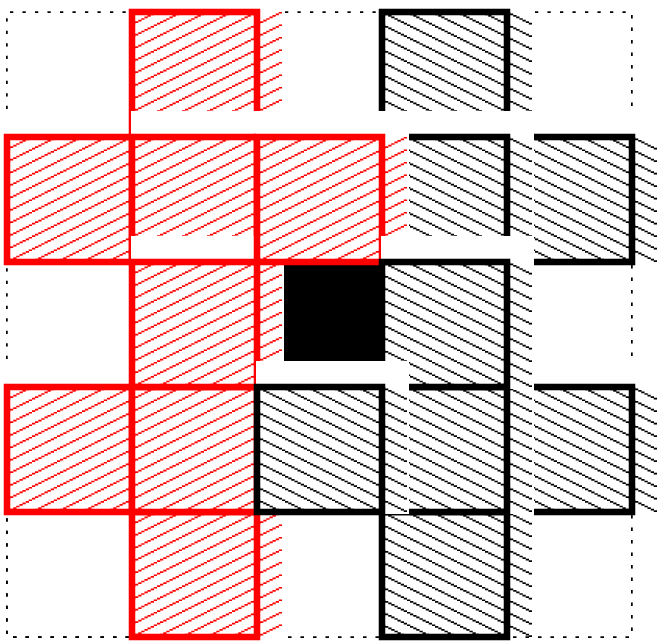}
\includegraphics[scale=0.5]{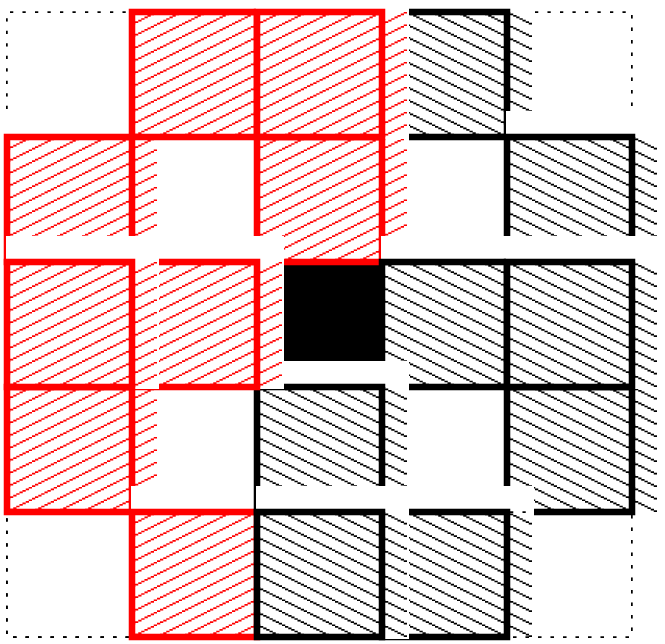}
\includegraphics[scale=0.5]{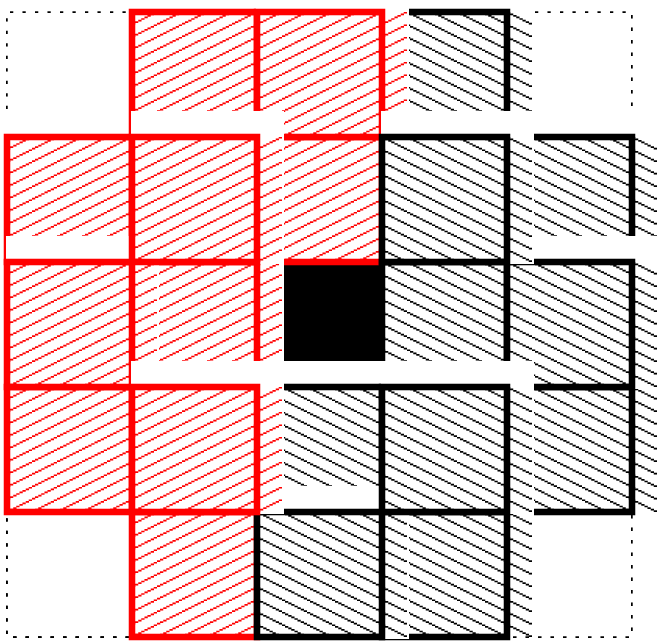}
\\[3mm]
\includegraphics[scale=0.5]{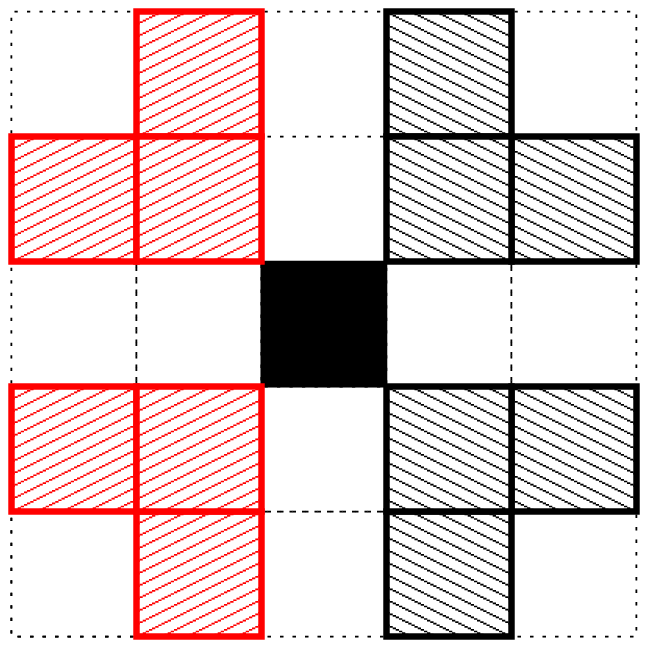}
\includegraphics[scale=0.5]{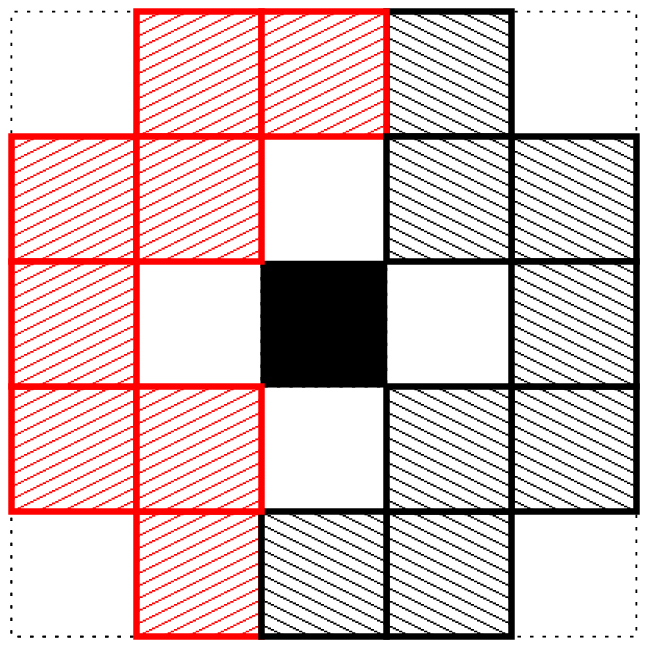}
\includegraphics[scale=0.5]{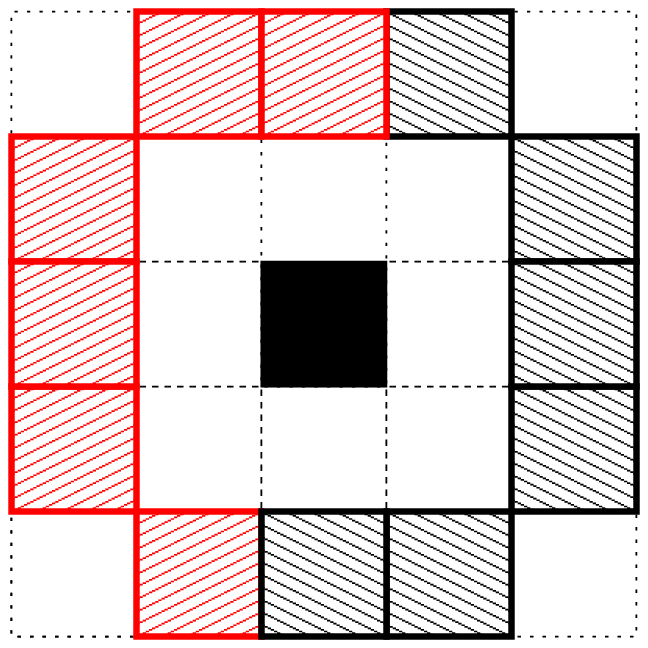}
\includegraphics[scale=0.5]{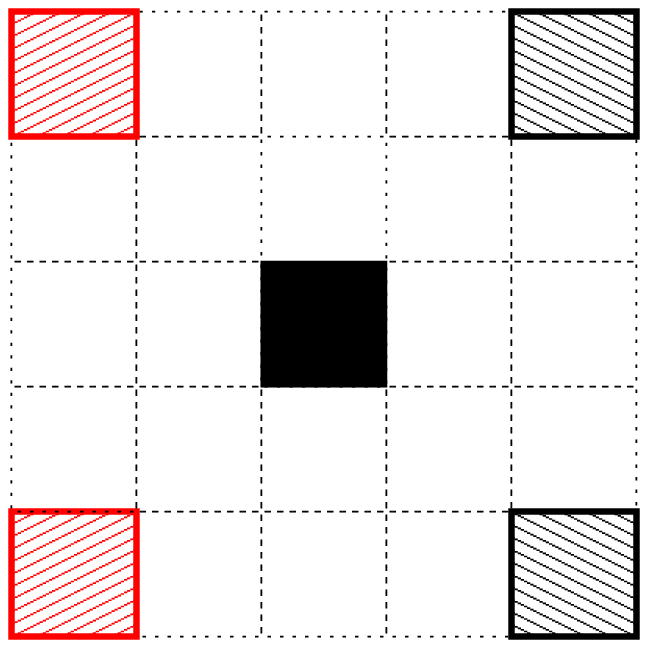}
\includegraphics[scale=0.5]{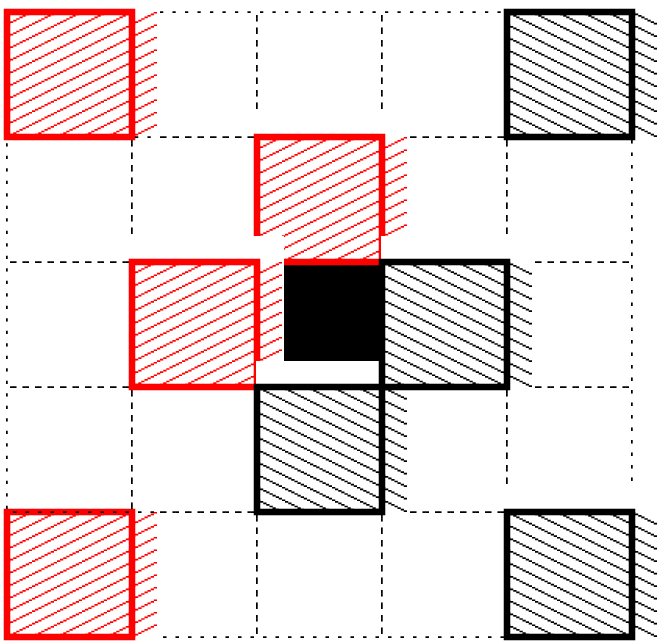}
\\[3mm]
\includegraphics[scale=0.5]{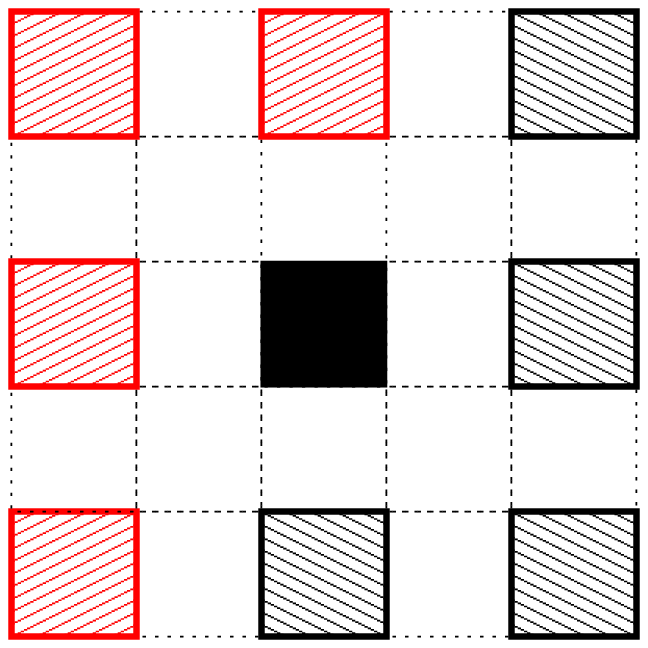}
\includegraphics[scale=0.5]{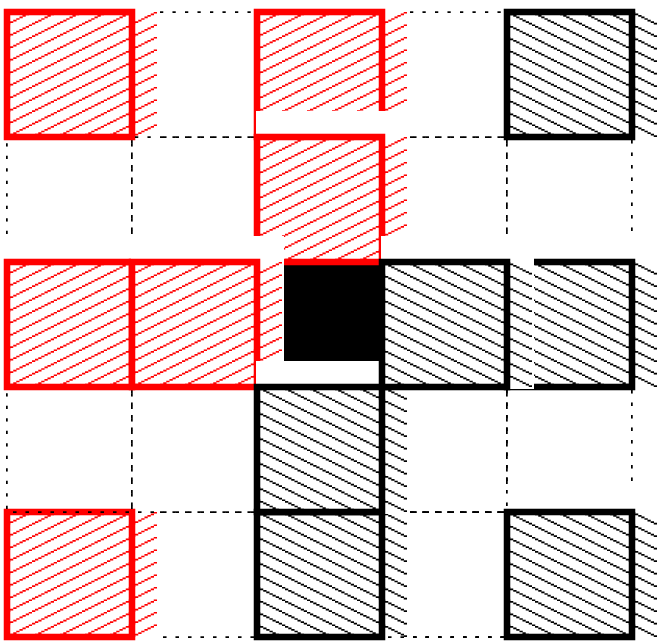}
\includegraphics[scale=0.5]{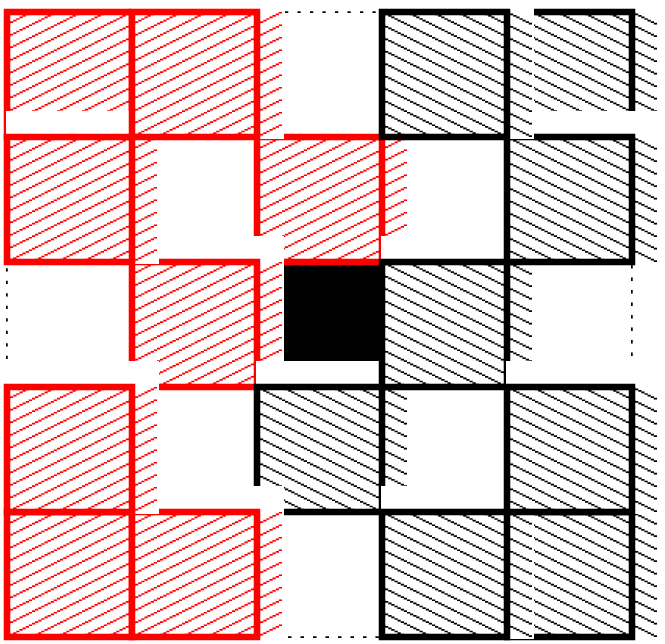}
\includegraphics[scale=0.5]{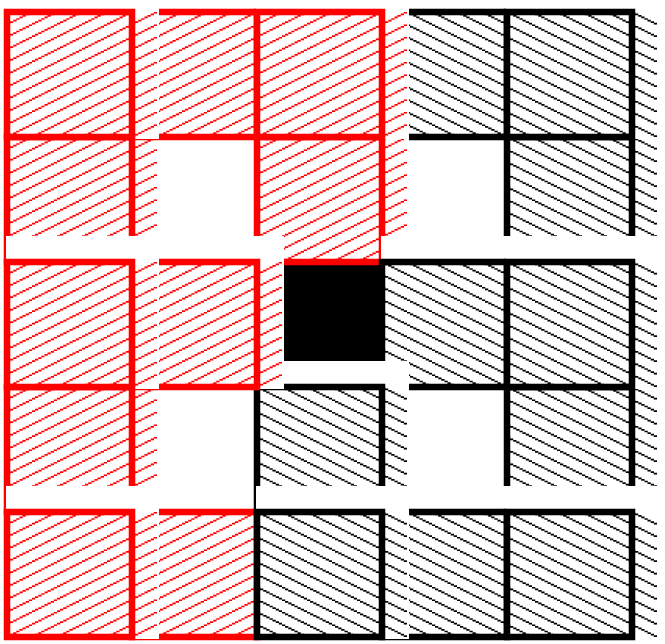}
\includegraphics[scale=0.5]{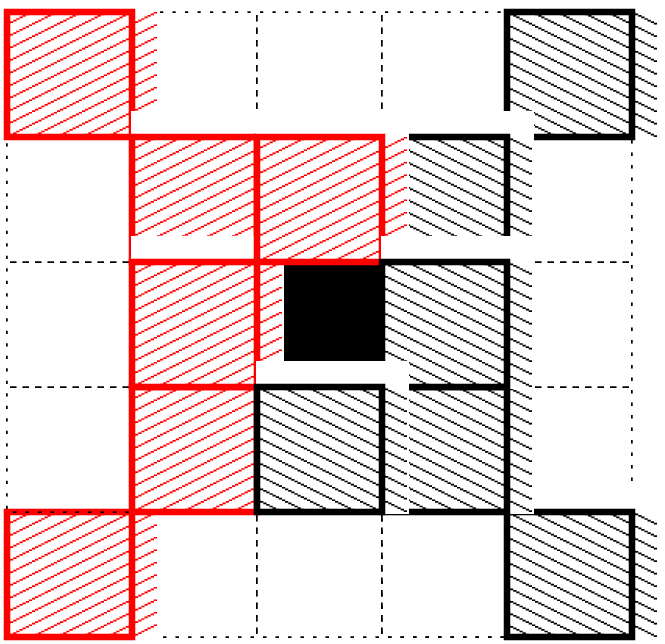}
\\[3mm]
\includegraphics[scale=0.5]{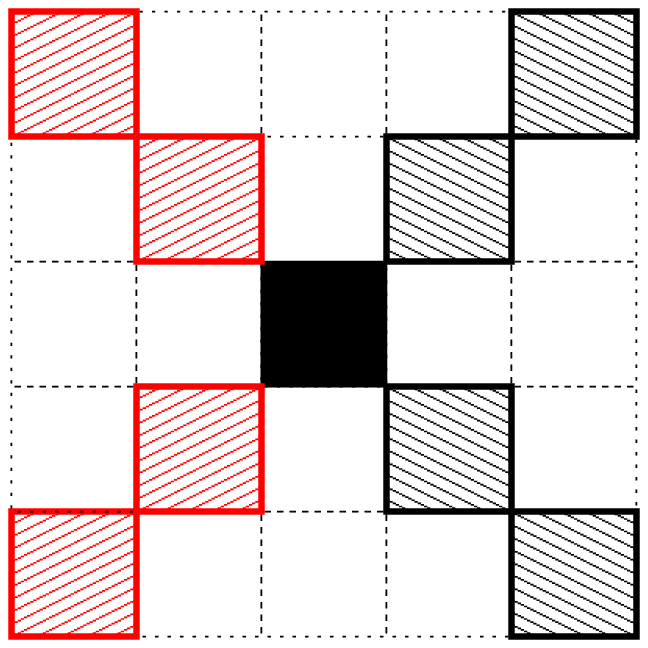}
\includegraphics[scale=0.5]{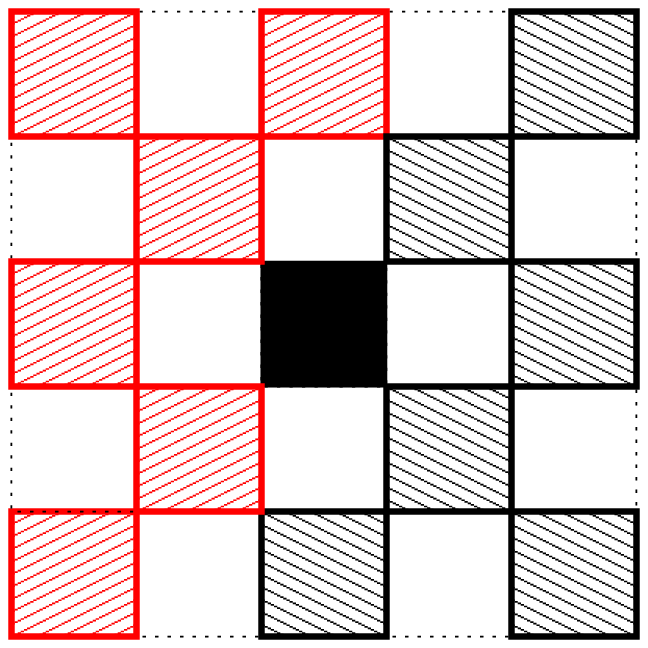}
\includegraphics[scale=0.5]{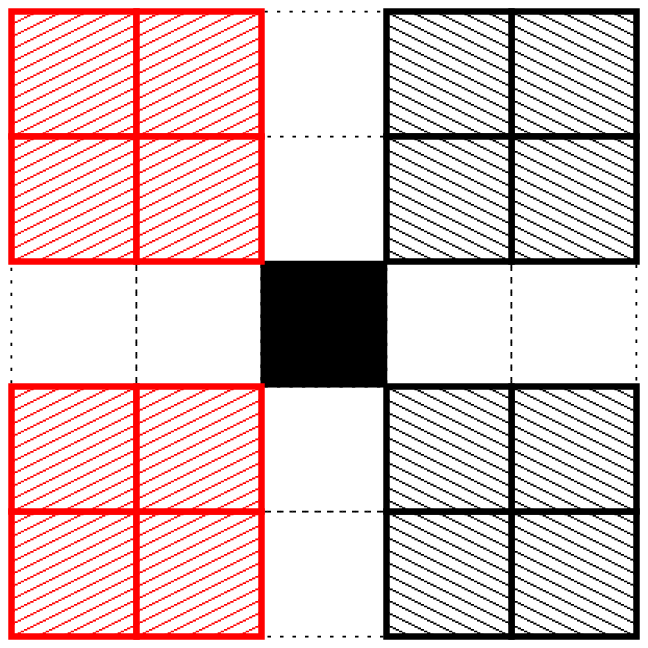}
\includegraphics[scale=0.5]{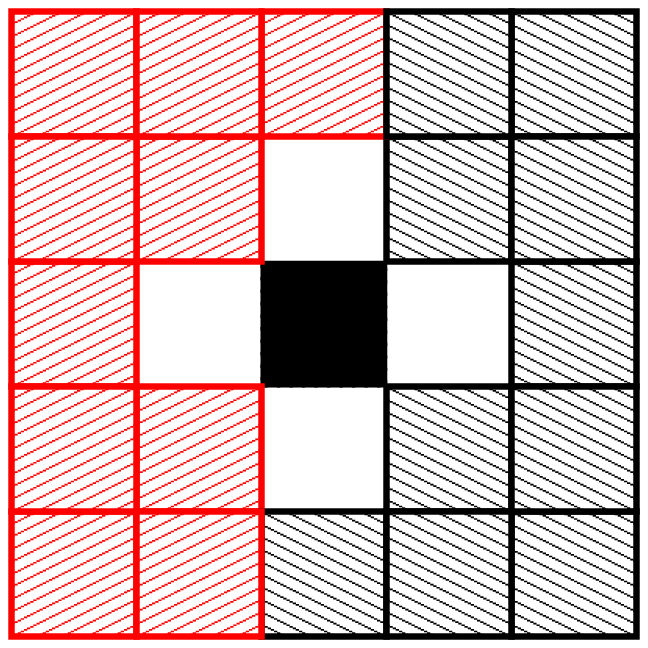}
\includegraphics[scale=0.5]{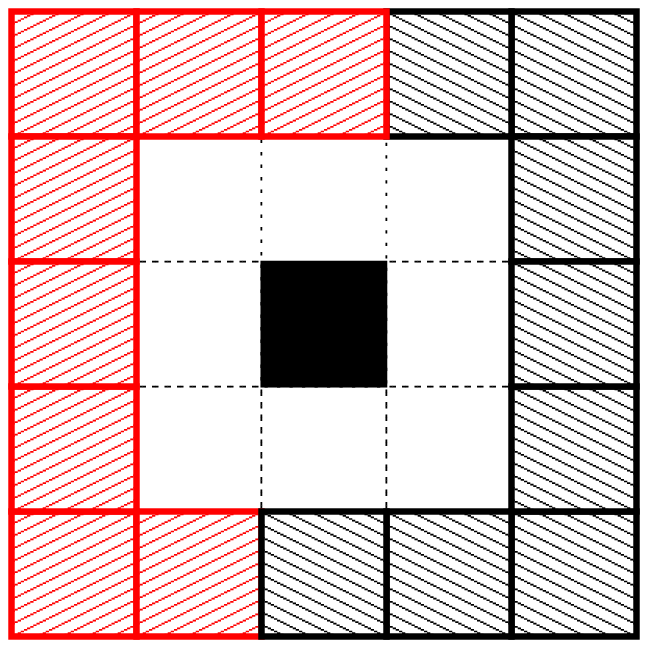}
\\[3mm]
\includegraphics[scale=0.5]{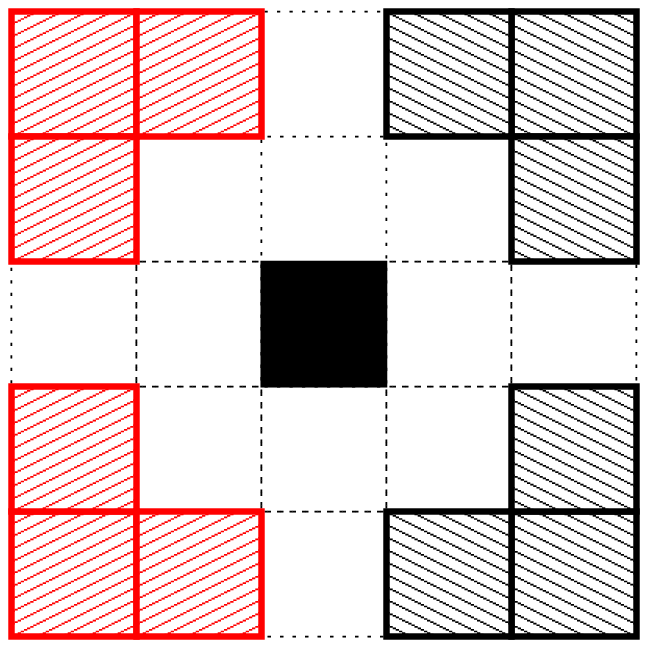}
\includegraphics[scale=0.5]{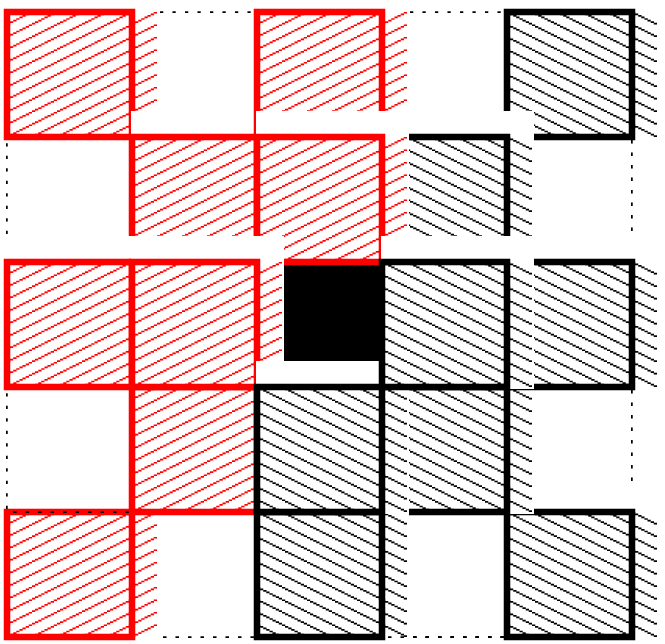}
\includegraphics[scale=0.5]{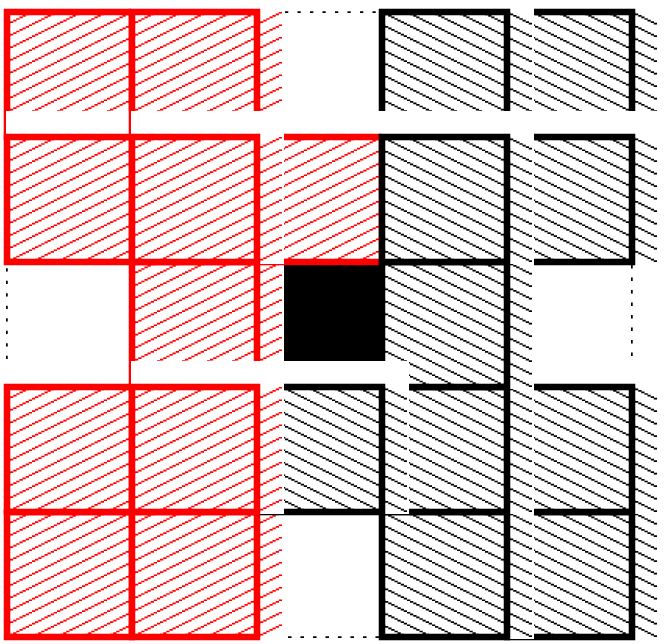}
\includegraphics[scale=0.5]{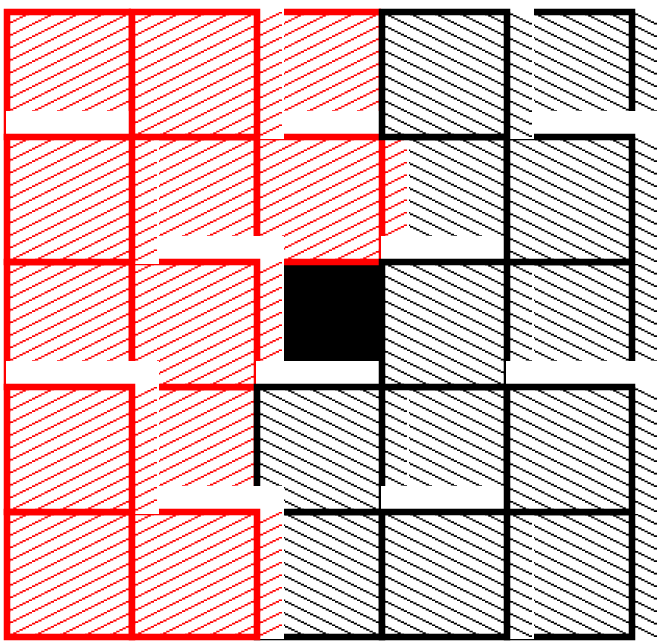}
\includegraphics[scale=0.5]{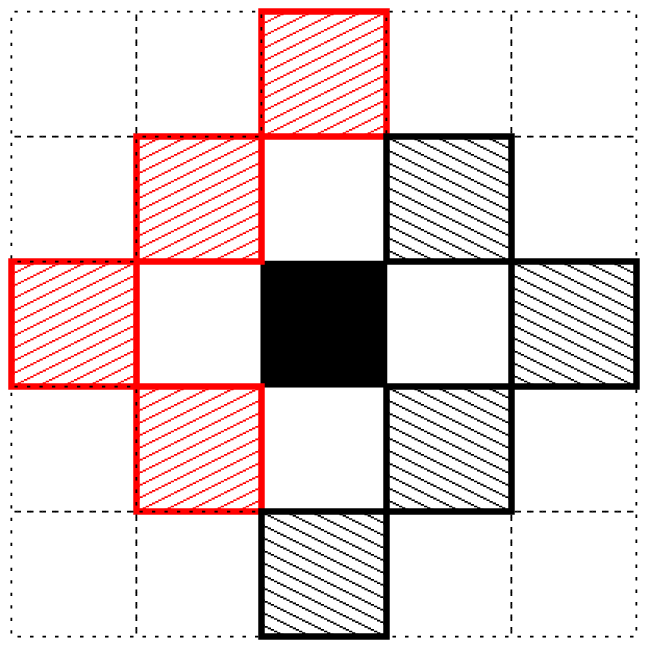}
\caption{All square lattice neighbourhoods $\mathcal{N}$ with 5N and 6N:
5N,	 5N+2N,	  5N+3N+2N,	   5N+4N+2N,	5N+4N+3N+2N,
5N+3N, 5N+4N+3N,	  5N+4N,	   6N,		6N+2N,
6N+4N, 6N+4N+2N,	  6N+5N+2N,	   6N+5N+4N+2N,	6N+3N+2N,
6N+3N, 6N+4N+2N,	  6N+5N+2N,	   6N+5N+4N+2N,	6N+5N+4N,
6N+5N, 6N+4N+3N+2N, 6N+5N+3N+2N, 6N+5N+4N+3N+2N.
Additionally the 4N+3N neighbourhoods is presented.
With the HKA on a square lattice when we assign the labels for the
investigated central site (black) we need to check already labelled and occupied sites in its neighbourhood (red slashed sites).
The possible links to remaining sites in the neighbourhood (black backslashed sites) may be checked later, basing on the neighbourhood's point symmetry.}
\label{fig-neighbours}
\end{figure*}

In this paper the square lattice random site percolation is reconsidered. 
With computer simulation we estimate the percolation thresholds for complex neighbourhoods $\mathcal{N}$ far beyond standard von Neumann's and Moore's ones.
We systematically examine neighbourhoods with radius $r=1$, $\sqrt{2}$, $2$, $\sqrt{5}$, $2\sqrt{2}$, i.e. which include the 4-th (5N, $r=2\sqrt{2}$) and the 5-th (6N, $r=\sqrt{5}$) nearest neighbours of sites.
These two neighbourhoods are combined with all neighbourhoods with smaller radius $r$, i.e. the next-next-nearest neighbours (4N, $r=2$), the next-nearest neighbours (3N, $r=\sqrt{2}$) and the  nearest neighbours (2N, $r=1$).
Some of these percolation thresholds were calculated earlier in Refs. \cite{pc-moore,km-sg}.
As in literature known to us also percolation threshold for 4N+3N neighbourhoods is missing we additionally compute its value as well.
The mentioned above neighbourhoods are presented in Fig. \ref{fig-neighbours}.

%% ----------------------------------------------------------------------------
%% ----------------------------------------------------------------------------

%% ############################################################################
\section{Calculations}
%% ############################################################################

\begin{figure}[!htbp]
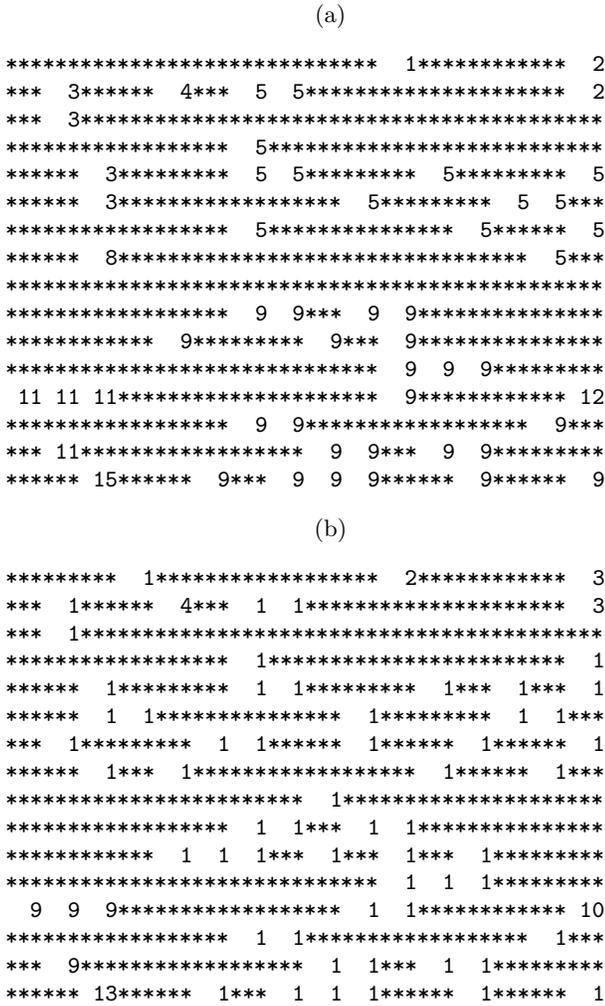

(a)
{\tt \small
\begin{verbatim}
******************************  1************  2
***  3******  4***  5  5*********************  2
***  3******************************************
******************  5***************************
******  3*********  5  5*********  5*********  5
******  3******************  5*********  5  5***
******************  5***************  5******  5
******  8*********************************  5***
************************************************
******************  9  9***  9  9***************
************  9*********  9***  9***************
******************************  9  9  9*********
 11 11 11*********************  9************ 12
******************  9  9******************  9***
*** 11******************  9  9***  9  9*********
****** 15******  9***  9  9  9******  9******  9
\end{verbatim}
}

(b)
{\tt \small
\begin{verbatim}
*********  1******************  2************  3
***  1******  4***  1  1*********************  3
***  1******************************************
******************  1************************  1
******  1*********  1  1*********  1***  1***  1
******  1  1***************  1*********  1  1***
***  1*********  1  1******  1******  1******  1
******  1***  1******************  1******  1***
************************  1*********************
******************  1  1***  1  1***************
************  1  1  1***  1***  1***  1*********
******************************  1  1  1*********
  9  9  9******************  1  1************ 10
******************  1  1******************  1***
***  9******************  1  1***  1  1*********
****** 13******  1***  1  1  1******  1******  1
\end{verbatim}
}
\caption{Example of labelled lattice with 5N+2N neighbourhood for (a) $p=0.20<p_c$ and (b) $p=0.25>p_c$.
Three stars {\tt ***} denote one empty site.}
\label{fig-hka}
\end{figure}

We use Hoshen--Kopelman algorithm \cite{hka} for the occupied sites labelling.
In the Hoshen--Kopelman scheme each site has one label: all sites in given cluster have the same labels and different clusters have assigned different labels.
The labelled lattice examples for 5N+2N neighbourhoods for the occupation probability $p$ below (a) and above (b) the percolation threshold $p_c$ are presented in Fig. \ref{fig-hka}.

%% ----------------------------------------------------------------------------
%% ----------------------------------------------------------------------------

The sigmoidal curve $P(p)$ tends to the Heaviside function $H(p-p_c)$ for $L\to\infty$.
The percolation threshold $p_c$ is evaluated numerically for a finite system as a crossing-point of the percolation probability $P$ dependence on sites occupation probability $p$ for various system sizes $L$ \cite{privman}.
Such strategy was applied successfully to investigate phase transition phenomena, including percolation \cite{km-sg,newman} or opinion dynamics \cite{stepper}.

Here, basing on $P(p)$ dependence for $L=100$, $500$ and $1000$ we look for the interval of the length $\Delta p=10^{-3}$ where these three curves cross each other.
For example for 2N case this interval is $(0.592,0.593)$.
These three significant figures of the left border of this interval estimate percolation threshold $p_c$ with three digits accuracy.

\begin{figure}[!htbp]
\psfrag{infty}{$\infty$}
\includegraphics[width=0.45\textwidth]{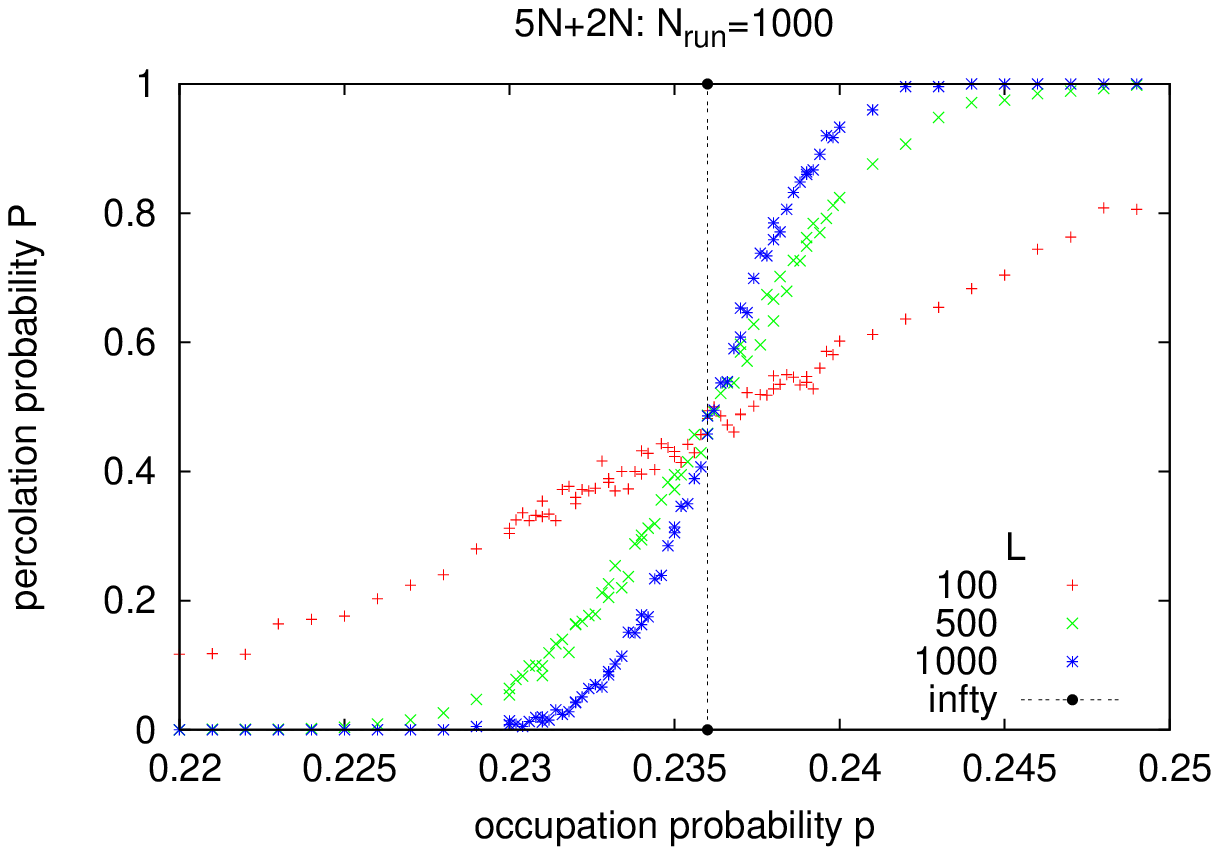}\\
\caption{Percolation probability $P$ vs. site occupation probability $p$ for 5N+2N neighbourhoods and three lattice sizes $L=100$, $500$ and $1000$.
The results are averaged out over $N_{\text{run}}=1000$ lattice realisations.
The common cross-point predicts percolation threshold $p_c$.}
\label{P-vs-p}
\end{figure}

In Fig. \ref{P-vs-p} examples of $P(p)$ dependence for 5N+2N neighbourhoods and for $L=100$, $500$ and  $1000$ are presented.

%% ----------------------------------------------------------------------------
%% ----------------------------------------------------------------------------

%% ############################################################################
\section{Results}
%% ############################################################################

%% ----------------------------------------------------------------------------
\begin{table}[!ht]
\caption{
Percolation threshold $p_c$ for square lattice site percolation and various neighbourhoods constructed with 5N an 6N.
}
\label{tab-1}
\begin{ruledtabular}
\begin{tabular}{rl l}
$z$ & $\mathcal{N}$       & $p_c$         \\
\hline
  4 & 2N, 3N, 4N, 6N      & $\theta$      \\
\hline
  8 & 3N+2N, 4N+3N, 6N+4N & $1-\theta$    \\
    & 4N+2N, 6N+3N        & $0.337\cdots$ \\
    & 5N                  & $0.270\cdots$ \\
    & 6N+2N               & $0.277\cdots$ \\
\hline
 12 & 4N+3N+2N, 6N+4N+3N  & $0.288\cdots$ \\
    & 5N+2N               & $0.236\cdots$ \\
    & 5N+3N               & $0.225\cdots$ \\
    & 5N+4N               & $0.221\cdots$ \\
    & 6N+3N+2N            & $0.240\cdots$ \\
    & 6N+4N+2N            & $0.233\cdots$ \\
    & 6N+5N               & $0.199\cdots$ \\
\hline
 16 & 5N+3N+2N            & $0.219\cdots$ \\
    & 5N+4N+2N            & $0.208\cdots$ \\
    & 5N+4N+3N            & $0.202\cdots$ \\
    & 6N+5N+2N            & $0.187\cdots$ \\
    & 6N+5N+3N            & $0.182\cdots$ \\
    & 6N+5N+4N            & $0.179\cdots$ \\
    & 6N+4N+3N+2N         & $0.208\cdots$ \\
\hline
 20 & 5N+4N+3N+2N         & $0.196\cdots$ \\
    & 6N+5N+3N+2N         & $0.177\cdots$ \\
    & 6N+5N+4N+2N         & $0.172\cdots$ \\
    & 6N+5N+4N+3N         & $0.167\cdots$ \\
\hline
 24 & 6N+5N+4N+3N+2N      & $0.164\cdots$ \\
\end{tabular}
\end{ruledtabular}
\end{table}
%% ----------------------------------------------------------------------------

\begin{figure}
\includegraphics[width=0.45\textwidth]{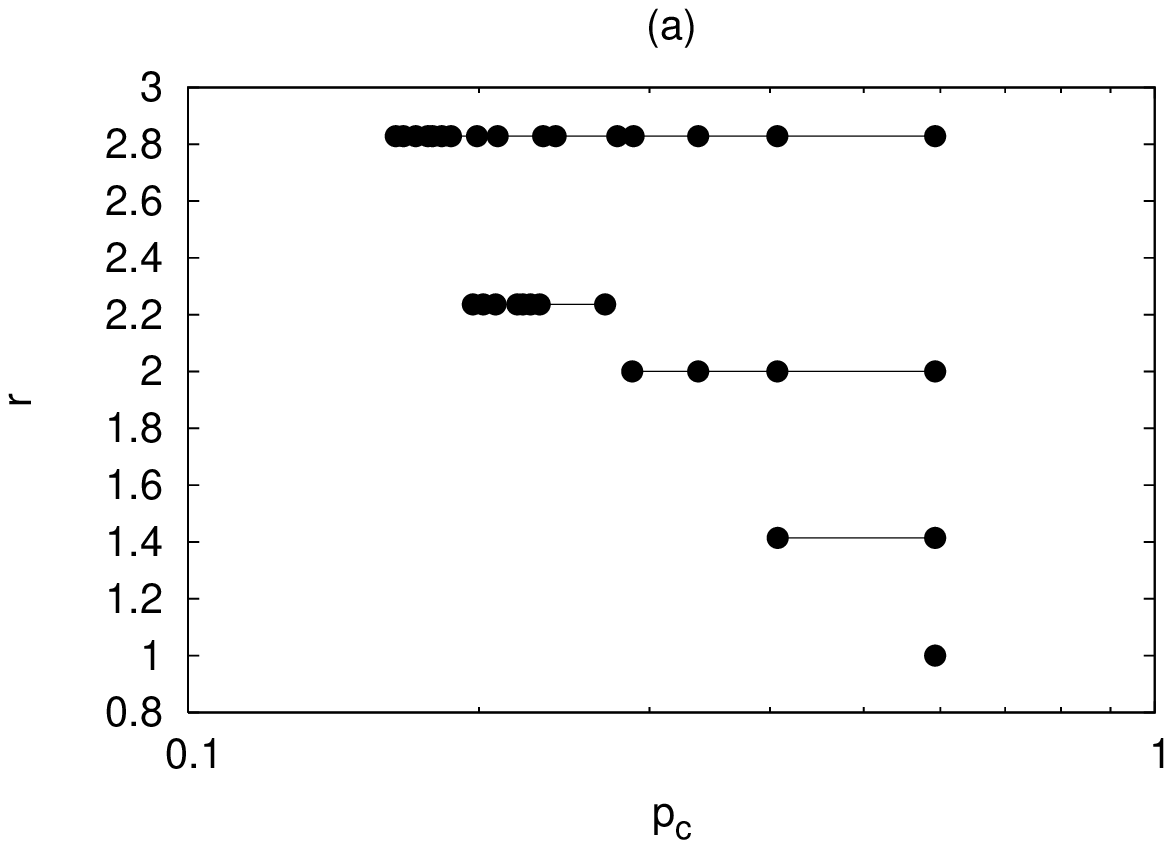}\\
\includegraphics[width=0.45\textwidth]{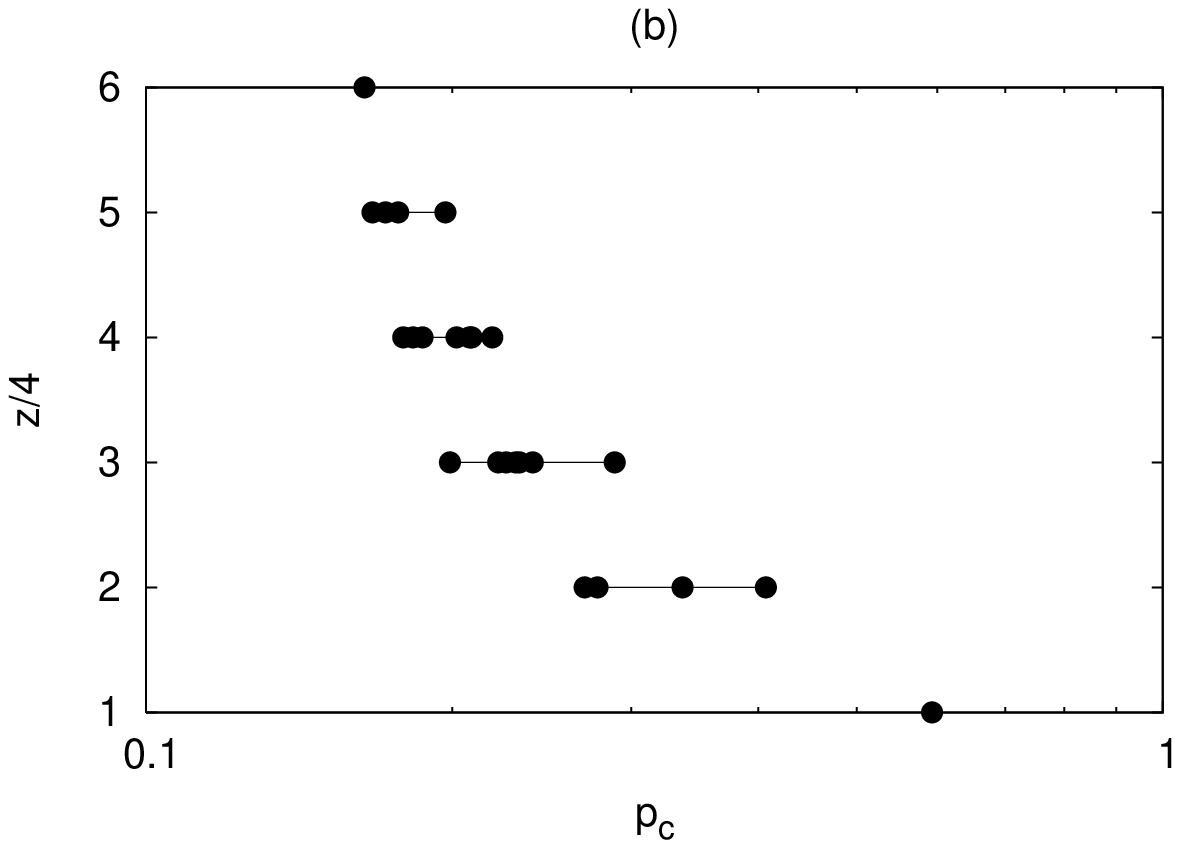}
\caption{Percolation threshold $p_c$ for various 
(a) neighbourhood radius $r$ and 
(b) coordination number $z$.}
\label{pc-vs-rz}
\end{figure}

The calculated percolation thresholds are collected in Tabs. \ref{tab-1} and \ref{tab-2}.
In Tab. \ref{tab-1} neighbourhoods are sorted according to increasing coordination number $z$ while in Tab. \ref{tab-2} according to neighbourhood radius $r$ (and lexicographically).
For completeness, in both tables, the results from Ref. \cite{km-sg} are included.
The results are also presented in Fig. \ref{pc-vs-rz}.

%% ----------------------------------------------------------------------------
%% ----------------------------------------------------------------------------

%% ----------------------------------------------------------------------------
\begin{table}[!ht]
\caption{Percolation threshold $p_c$ for square lattice site percolation and various neighbourhoods $\mathcal{N}$ sorted lexicographically, with decreasing neighbourhoods radius $r$.}
\label{tab-2}
\begin{ruledtabular}
\begin{tabular}{rr lll}
$r$         & $z$ & $\mathcal{N}$  & $p_c$ & Ref.\\
\hline
$2\sqrt{2}$ & 24 & 6N+5N+4N+3N+2N & $0.164\cdots$ & --- \\
            & 20 & 6N+5N+4N+3N    & $0.167\cdots$ & --- \\
            & 20 & 6N+5N+4N+2N    & $0.172\cdots$ & --- \\
            & 16 & 6N+5N+4N       & $0.179\cdots$ & --- \\
            & 20 & 6N+5N+3N+2N    & $0.177\cdots$ & --- \\
            & 16 & 6N+5N+3N       & $0.182\cdots$ & --- \\
            & 16 & 6N+5N+2N       & $0.187\cdots$ & --- \\
            & 12 & 6N+5N          & $0.199\cdots$ & --- \\
            & 16 & 6N+4N+3N+2N    & $0.208\cdots$ & --- \\
            & 12 & 6N+4N+3N       & $0.288\cdots$ & --- \\
            & 12 & 6N+4N+2N       & $0.233\cdots$ & --- \\
            &  8 & 6N+4N          & $1-\theta$ & --- \\
            & 12 & 6N+3N+2N       & $0.240\cdots$ & --- \\
            &  8 & 6N+3N          & $0.337\cdots$ & --- \\
            &  8 & 6N+2N          & $0.277\cdots$ & --- \\
            &  4 & 6N             & $\theta$ & \cite{km-sg} \\
\hline
$\sqrt{5}$  & 20 & 5N+4N+3N+2N    & $0.196\cdots$ & --- \\
            & 16 & 5N+4N+3N       & $0.202\cdots$ & --- \\
            & 16 & 5N+4N+2N       & $0.208\cdots$ & --- \\
            & 12 & 5N+4N          & $0.221\cdots$ & --- \\
            & 16 & 5N+3N+2N       & $0.219\cdots$ & --- \\
            & 12 & 5N+3N          & $0.225\cdots$ & --- \\
            & 12 & 5N+2N          & $0.236\cdots$ & \cite{km-sg} \\
            &  8 & 5N             & $0.270\cdots$ & --- \\
\hline
2           & 12 & 4N+3N+2N       & $0.288\cdots$ & \cite{km-sg} \\
            &  8 & 4N+3N          & $1-\theta$ & --- \\
            &  8 & 4N+2N          & $0.337\cdots$ & \cite{km-sg} \\
            &  4 & 4N             & $\theta$ & \cite{km-sg} \\
\hline
$\sqrt{2}$  &  8 & 3N+2N          & $1-\theta$ & \cite{pc-moore} \\
            &  4 & 3N             & $\theta$ & \cite{km-sg} \\
\hline
$1$         &  4 & 2N             & $\theta$ & \cite{pc-sl} \\
\end{tabular}
\end{ruledtabular}
\end{table}
%% ----------------------------------------------------------------------------

%% ############################################################################
\section{Conclusions}
%% ############################################################################

Concluding, we have presented the random site percolation thresholds $p_c$ for square lattice with complex neighbourhoods which is constituted by 4-th and 5-th nearest neighbours.
Single site in such neighbourhoods has from $z=4$ to $z=24$ neighbours at distances from $r=1$ to $r=2\sqrt 2$ from the neighbourhood centre.
For fixed values of either $z$ or $r$ the percolation threshold form a wide interval of values.
These ranges overlap each other supply earlier observations that $p_c$ have not to be decreasing function of coordination number $z$ \cite{wierman}.
We show the same for neighbourhood radius $r$.

As it was stated in Ref. \cite{km-sg} $p_c(\text{6N})=p_c(\text{4N})=p_c(\text{3N})=p_c(\text{2N})=\theta$.
Geometrical arguments similar to those presented in Ref. \cite{km-sg} yield 
\begin{subequations}
\begin{equation}
 p_c(\text{6N+4N})=p_c(\text{4N+3N})=p_c(\text{3N+2N}),
\end{equation}
\begin{equation}
 p_c(\text{4N+2N})=p_c(\text{6N+3N})
\end{equation}
 and 
\begin{equation}
 p_c(\text{6N+4N+3N})=p_c(\text{4N+3N+2N}).
\end{equation}
\end{subequations}
We improve value of $p_c(\text{5N})=0.270$, as well.

Our results may support further searching of universal formula for $p_c$ dependence on space dimension $d$, site coordination number $z$ and additionally the neighbourhoods radius $r$.
However --- as we can see from Tab. \ref{tab-2} --- these three numbers may be still insufficient to build the universal formula \cite{pc-sahimi,pc-GM-1,pc-GM-2} as several lattices with different neighbourhoods has exactly the same $d$, $z$ and $r$
but different $p_c$.

%% ============================================================================
\begin{acknowledgments}
The numerical calculations were carried out in ACK\---CY\-F\-RO\-NET\---AGH.
The machine time on HP Integrity Superdome is financed by the Polish Ministry of Science and Higher Education under grant No. MNiI/\-HP\_I\_SD\-AGH/\-047/\-2004.
\end{acknowledgments}
%% ============================================================================

%% ############################################################################

\end{document}